\pdfoutput=1
\documentclass[twocolumn,showpacs,preprintnumbers,amsmath,amssymb]{revtex4}

\usepackage{graphicx}
\usepackage{dcolumn}
\usepackage{bm}

\begin{document}
\newcommand{\mf}[1]{\boldsymbol{#1}}


\title{High Harmonic Generation via Continuum Wave-Packet Interference}
\author{Markus~C.~Kohler}
\author{Christian~Ott}
\author{Philipp~Raith}
\author{Robert~Heck}
\author{Iris~Schlegel}
\author{Christoph~H.~Keitel}
\author{Thomas~Pfeifer}
\email{tpfeifer@mpi-hd.mpg.de}
\affiliation{Max-Planck-Institut f\"ur Kernphysik, Saupfercheckweg 1, 69117 Heidelberg, Germany}

\date{\today}

\begin{abstract}
High-order harmonic generation (HHG) is investigated theoretically in the over-the-barrier ionization (OBI) regime revealing
the strong signature of interference between two separately ionized and separately propagating free wave packets of a single electron. The interference leads to the emission of coherent light at a photon energy corresponding to the kinetic-energy difference of the two recolliding electron quantum paths, thus complementary to the well-known classical three-step picture of HHG.
As will be shown by time-frequency analysis of the emitted radiation, the process entirely dominates the coherent HHG emission after the atomic ground state has been depleted by a strong field. Moreover, it can be isolated from the continuum--bound harmonics via phase-matching.
\end{abstract}

                             

                             
\maketitle
High-order harmonic generation (HHG) is a key process in ultrafast science. As a fundamental example of strong-field electron dynamics, it is fascinating both from a fundamental physics perspective --- coherent recollision of sub-femtosecond ionized electronic wavefunction within a fraction of an optical cycle of the intense driving laser~\cite{CORKUM1993,LEWENSTEIN1994,KULANDER1993,PUKHOV2003} --- but also from a technological point of view enabling the creation of attosecond-duration flashes of light~\cite{PAUL2001,HENTSCHEL2001} for the time-resolved study of the fastest processes in physics and chemistry (for current reviews see~\cite{CORKUM2007,PFEIFER2008}).  HHG is currently well-understood in terms of the three-step model~\cite{CORKUM1993,KULANDER1993}: An electronic wave packet is partially ionized, accelerated in the laser field, and returns to its parent ion.  At that point, the ionized and bound-state portions of the electronic wave packet interfere giving rise to a strong, coherent high-frequency dipole response that can lead to the emission of an HHG photon along with the recombination of the electron into the bound state.   
The coherent HHG emission from this continuum--bound (CB) mechanism is practically absent as soon as the atom is fully ionized by the laser pulse~\cite{PUKHOV2003}. Then, recombination to the ground state can only proceed by incoherent spontaneous emission of radiation that scales linearly with the number of atoms $N$ in the gas target rather than $\propto N^2$ as in the case of the very efficient typically considered coherent process.

However, the impact of free portions of the electronic wave packet alone has not been fully recognized yet.  
 Continuum--continuum (CC) transitions in atomic HHG have been studied with focus on the Bremsstrahlung emission of single wave packets recolliding with a bare core \cite{EMELIN2008} and  the interaction with a continuum wave packet ionized just shortly  before the recollision time  \cite{MILOSEVIC2003,PLAJA2007}.
Also, in a recent theoretical work, it was shown that recollisions of two free electrons after strong-field double-ionization can lead to weak HHG emission beyond the usual cutoff~\cite{KOVAL2007}.  

Here, we present a novel paradigm and generalization of atomic HHG by predicting the efficient production of coherent soft-x-ray light from the interference of spectrally distinct free \emph{single-electron} wave packets.  It is shown both analytically and numerically that even for the case of a fully depleted single-electron atomic system (no bound states populated) a characteristic coherent polarization response can be obtained corresponding to the energy difference of two distinct continuum electronic states. Investigating the temporal structure of emission reveals that the process dominates the traditionally known CB interference of the electron recombining with the bound states at times after the ground-state population has been depleted.

In most general terms, the continuum--continuum (CC) interference mechanism
presented here provides a fundamental physical picture: wave function splitting or spreading, subsequent simultaneous recollision with different energies, and core-mediated transition with photoemission at the
difference energy.  This approach thus generalizes the traditional picture
of CB HHG to include CC transitions.

We start out by describing the key idea considering the case of an electron--atom collision.  The coherent part of the emitted radiation can be calculated via the expectation value of the acceleration using Ehrenfest's theorem:
\begin{equation}
 \mf{a}(t)=-\langle \Psi(t)\vert \nabla V\vert \Psi(t)\rangle \label{a_general}
\end{equation}
with the ionic potential $V$ and atomic units {\it(a.u.)} being used throughout.
Note that this approach, in consistence with earlier and current atomic HHG theory, considers only the coherent part of the emission response and neglects incoherent spontaneous emission which cannot be phase-matched.
We briefly describe the general conditions required for the emission of radiation: If $\vert \Psi(t)\rangle$ is an eigenstate of the Hamiltonian, the expectation value is time-independent. Harmonic emission can only occur for a linear combination of at least two eigenstates:  $\vert\Psi(t)\rangle=a_{1} \vert \epsilon_1\rangle e^{-i \epsilon_1 t}+a_{2} \vert \epsilon_2\rangle e^{-i \epsilon_2 t}$. Separating the time-dependent part $a(t)=-a_1 a_2 \langle \epsilon_1\vert \nabla V\vert\epsilon_2\rangle e^{-i (\epsilon_2-\epsilon_1) t}+c.c.$ shows the resulting dipole oscillates with the difference energy between the two states. CB HHG can be understood in this context: A recolliding electronic plane wave (energy $\epsilon_1$) interferes with the bound wave packet (energy $-I_p$) and the difference energy ($\epsilon_1+ I_p$) is emitted.
Moreover, if a bichromatic electron wave sweeps over the potential, it would lead to light emission at the kinetic energy difference. This intuitive model strictly holds only for time-independent Hamiltonians but its picture remains in atomic HHG with collision times typically short on the time scale of a laser period.

\begin{figure}
\centering
\includegraphics[width=\linewidth]{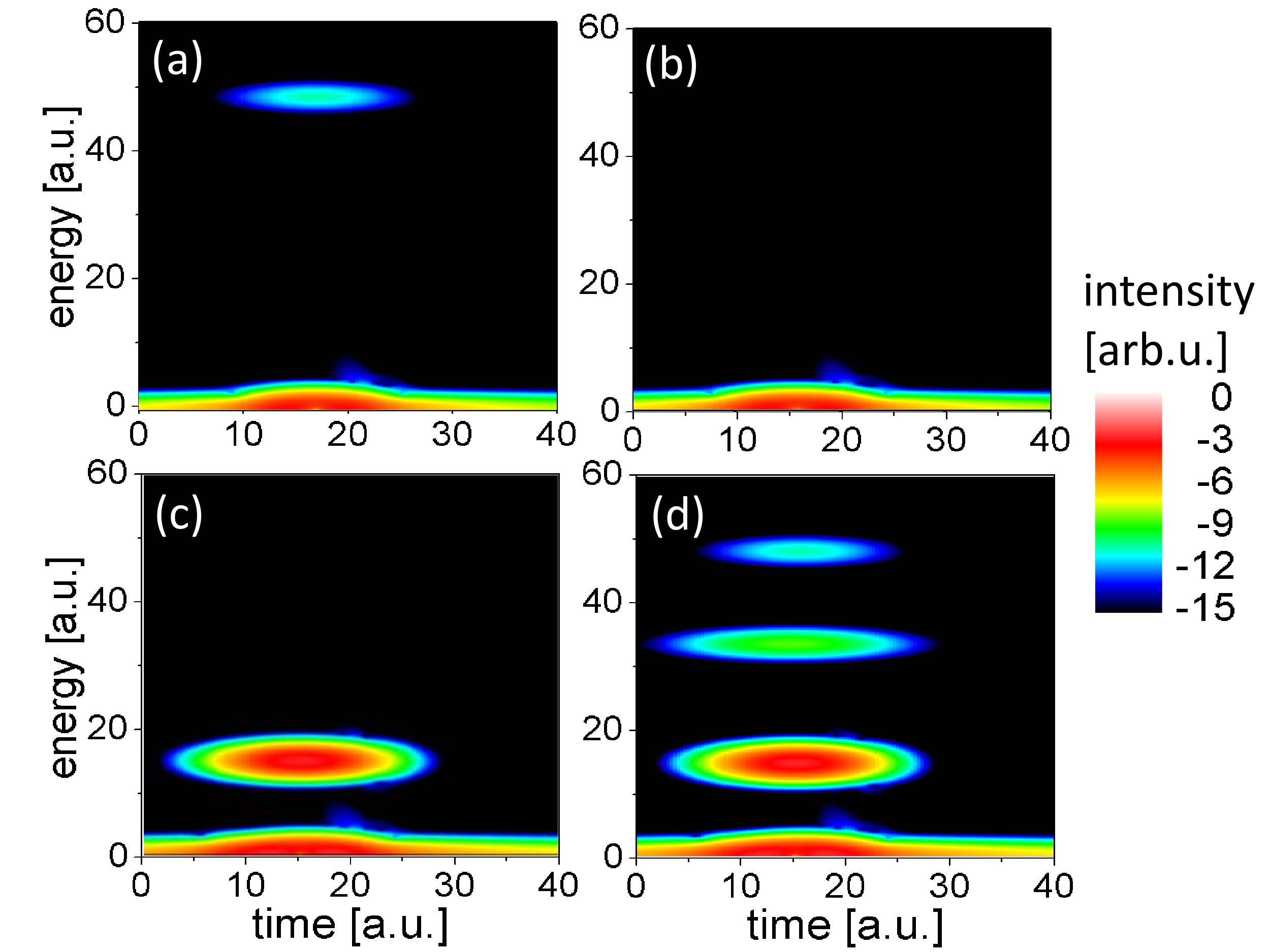}
\caption{\label{fig-collision}
(color online) Electron--atom collision and analysis of the dipole acceleration response.  Windowed Fourier transforms on a logarithmic scale are shown for four cases: a) monochromatic electron wave packet (47.8~a.u. kinetic energy) colliding with a partially ionized atom, b) same as a) but for a fully ionized atom, c) a bichromatic electron wave packet (47.8~a.u., 33.1~a.u. kinetic energies) colliding with a fully ionized atom, d) same as c) but for a partially ionized atom.  In c) and d) the signature of cc electron interference is observed at the difference kinetic energy of 14.7~a.u.}
\end{figure}

 To numerically illustrate this difference-frequency mechanism, we solve the time-dependent Schr\"odinger equation for a 1-dimensional hydrogen atom using a smoothed Coulomb potential $V(x)=1/\sqrt{x^2+a^2}$  with the smoothing parameter $a=1.4039$ chosen to match the hydrogen ionization potential.  We consider 4 cases: a) a monochromatic free-electron wave packet colliding with a partially ionized atom (the conventional HHG scenario), b) the same as a) but for a fully ionized atom, c) a bichromatic free electron colliding with a fully ionized atom, and d) a bichromatic free-electron wave packet colliding with a partially ionized atom.  We analyze the dipole acceleration expectation value by means of a windowed Fourier transform (Gabor transform) with a temporal window size of 3~a.u.  The results are shown in Fig.~\ref{fig-collision}.

In case a), we observe the dipole response at a frequency corresponding to the sum of electron kinetic energy and ionization potential of the atom $E_\mathrm{kin}+I_\mathrm{p}$, well-known from traditional HHG.  In case b), no high-energy dipole response can be observed.  The low energy signal corresponds to the Bremsstrahlung spectrum \cite{EMELIN2008} arising due to its non-vanishing spectral width and in fact can be understood as a special case of the mechanism discussed here.  Case c) exhibits a dipole response at a slightly lower frequency than the one obtained in a), while case d) shows a total of 3 different frequencies.  The absence of any high-energy emission in case b) arises from the lack of interference as only one electronic state takes place in the interaction in consistence with an earlier theoretical study~\cite{PUKHOV2003}.  Emission in c) is created by the interference of two free-electron wave packets with kinetic energies of $E_\mathrm{kin,1}$ and $E_\mathrm{kin,2}$ that create a time-dependent dipole moment corresponding to a beating frequency of the difference $E_\mathrm{kin,1}-E_\mathrm{kin,2}$ inside the non-linear potential set up by the Coulomb field of the atom.  This is the origin of the coherent CC
 electron emission process for HHG described in more detail below.  Case d) shows emission due to both free-electron wave packets, each interfering with the bound state and amongst themselves, giving rise to three distinct dipole oscillation frequencies.  Note that CC transitions also occur when the two wave packets meet each other in the potential with opposite (different absolute) momenta but with a lower probability as a consequence of the higher momentum transfer required.

\begin{figure}
\centering
\includegraphics[width=\linewidth]{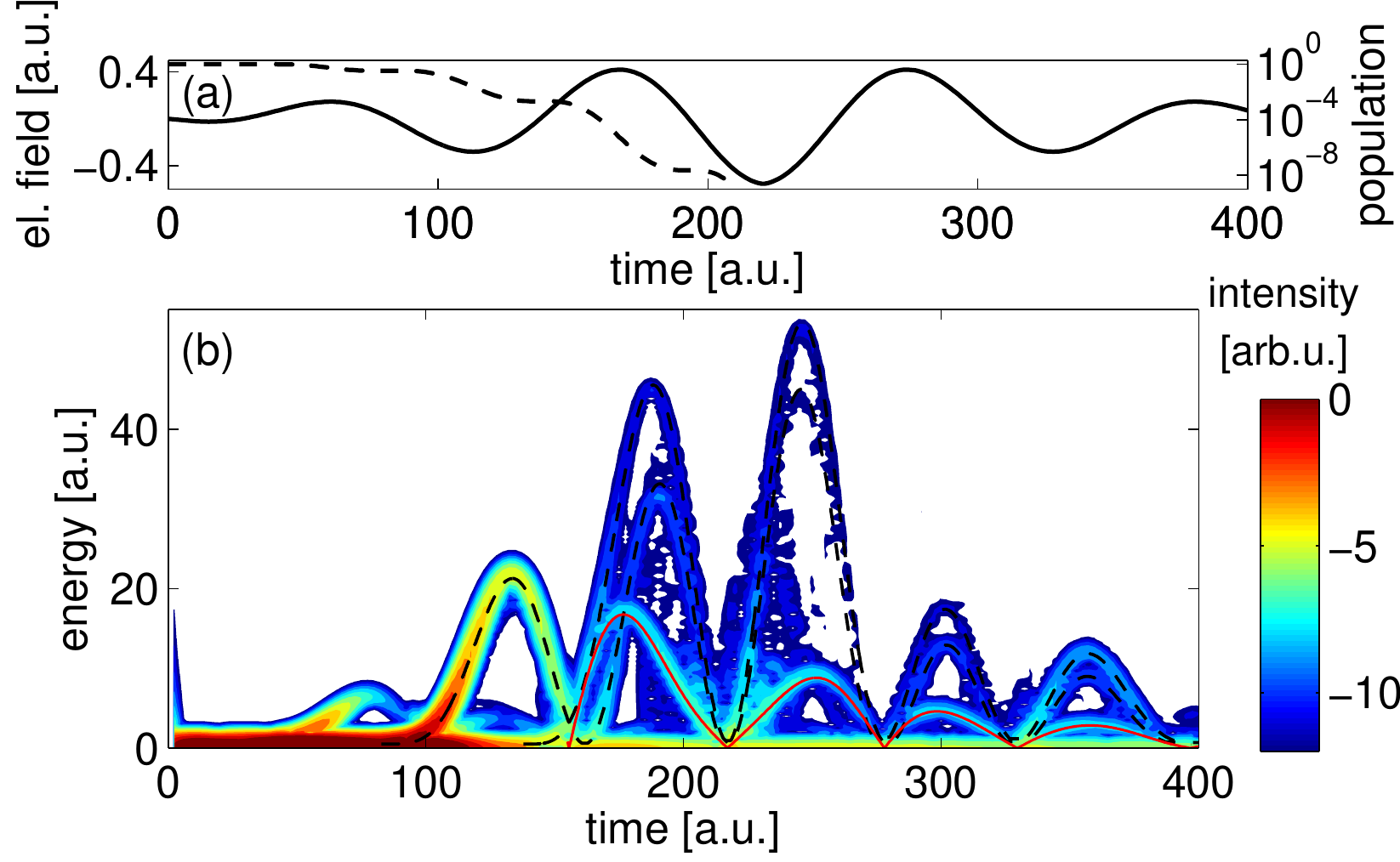}
\caption{\label{fig-HHGnumerical}
(color online) Time-frequency analysis of HHG showing the signature of CC wave-packet interference.  a) Driver pulse used in the simulation (solid line) and the reliable part of the ground-state population (dashed line) calculated via \cite{ionrate}.  b) Windowed Fourier transform of the HHG emission on a logarithmic scale.  Dashed black lines: classically calculated kinetic energies of electrons returning to the ion.  Solid red line: difference between the two black dashed curves.}
\end{figure}

In the following, it is shown that HHG radiation contains a signature of CC electron wave packet interference when approaching the over-the-barrier (OBI) regime.  As a more realistic model, in this case, we use a hydrogen atom with $V(r)=-1/r$ and solve the 3-dimensional time-dependent Schr\"odinger equation in a strong field but in principle, any other atom, ion or molecule could be chosen. 
The peak intensity of $10^{16}$W/cm$^2$ of the laser pulse (shown in Fig.~\ref{fig-HHGnumerical}a) is chosen such that almost complete depletion of the ground state occurs on the leading edge of the pulse.  The evolution of the ground state population is shown in Fig.~\ref{fig-HHGnumerical}a) via $n_0(t)=\exp{(-\int_{-\infty}^td\tau w(\tau))}$ where $w(\tau)$ is the ionization rate being a combination of the empirical formulas in Ref.~\cite{ionrate}. This way, CC transitions become the dominating HHG mechanism for large parts of the pulse. The laser intensity and recollision energies are higher than in most of the current HHG experiments but our results are not specific to these intensities.
For solving the TDSE, the code QPROP~\cite{BAUER2006} is used which expands the wave function in spherical harmonics and propagates them within a Crank--Nicolson scheme. As simulation parameters, we used a time step of $\Delta t=0.01$ a.u., a radial grid of $r_{max}=800$~a.u. with a spacing of $\Delta r=0.1$~a.u. and a highest angular momentum of $L=140$.   Again, we calculate the windowed Fourier transform of the dipole-acceleration response shown in Fig.~\ref{fig-HHGnumerical}b). 

To analyze the individual time-resolved frequency components contained in the dipole response, we use Newton's equations to calculate classical electron trajectories in the laser field originating from the ion.  
In Fig.~\ref{fig-HHGnumerical}b, dashed black line, we show the classical continuum--bound (recombination) energy of $E_\mathrm{kin,1,2}+I_\mathrm{p}$ as well as the energy difference $E_\mathrm{kin,1}-E_\mathrm{kin,2}$ of the two returning classical trajectories (solid red), each vs. the time of recollision. 
We compare those to the dipole-response obtained from our quantum calculation:
Until time $t=150$~a.u., HHG emission is caused by CB transitions that are still rather strong because the bound state has not been fully depleted (compare Fig.~\ref{fig-HHGnumerical} a). After time $t=150$~a.u., we can clearly observe both the signature of ground-state recombination (CB) as well as CC interference by comparing to the classically calculated return energies.  Interestingly, the CC component of the dipole response is the dominant contribution for several half-cycles between $t=150$~a.u. and $t=320$~a.u. This can be understood from the fact that depletion of the ground state occurs around $t=150$~a.u. and thus, interference with and recombination to the ground state can no longer take place. Instead, the two parts of the wave packet which have been ionized within the first two half cycles steadily interfere with each other in the atomic potential for the remainder of the laser pulse.  
 Note that the electronic interference described here is different from the optical interference of the harmonic field caused by the well-known short and long trajectories.

The significance of the signal is pointed out by estimating the photon yield in a small energy window between 12~a.u. and 18~a.u. Assuming a phase-matched gas volume with radius $100~\mu$m and length $1$~mm at a density of $10^{18}$/cm$^3$ and emitting into a solid angle of $10^{-9}$, we calculate a yield of $10^3$ photons per shot within the time window of $t=110$~a.u. and $t=155$~a.u. originating from CB transitions. Between $t=155$~a.u. and $t=200$~a.u., a photon number of $10^{-1}$  is emitted within this bandwidth mainly caused by CC transitions. The typical kHz repetition rates, therefore, allow for a measurement of this effect  that could potentially be enhanced by using larger atoms/molecules with stronger core potentials.

In the last part of this letter, we describe the CC recollision process and harmonic emission by an analytical theory framework based on the strong-field approximation (SFA).  
The standard approach to HHG within the SFA where the dipole moment is calculated \cite{LEWENSTEIN1994} is not sufficient since two plane continuum waves with different momenta have a vanishing dipole moment. As in the SFA no distortion of the plane waves by the potential of the atom is included, the impact of the potential is imposed by directly calculating the nonlinear dipole acceleration responsible for HHG emission via the expectation value of the acceleration (Eq.~\ref{a_general}). This way, the interaction of the electron with the core is included which is needed for the momentum conservation of the CC transition \cite{GORDON1995}.
Within the dipole approximation (DA), the electronic wavefunction can be written as
\begin{eqnarray}
 \vert\Psi(t)\rangle&=&\vert \phi_B (t)\rangle+\vert \phi_C (t)\rangle \label{wf} \\ \nonumber
&=&\sqrt{n_0(t)}\vert0\rangle e^{i I_p t}-i \int d^3 p \vert \mf{p}+\mf{A}(t) \rangle \times \\ \nonumber 
\times &&\!\!\!\!\!\!\!\!\!\!\!\!\int_{-\infty}^t\!\!\!\!\!\! dt'  \sqrt{n_0(t')}\langle \mf{p}+\mf{A}(t')\vert \mf{r}\cdot \mf{E}(t')\vert 0\rangle e^{-i(S(\mf{p},t,t')-I_p t')}, 
\end{eqnarray}
where $S(\mf{p},t,t')=\int_{t'}^{t}d\tau(\mf{p}+\mf{A}(\tau))^2/2$ is the classical action.
Inserting \eqref{wf} into \eqref{a_general} yields two contributing terms~\cite{PLAJA2007}: The CB transition can be identified as $\mf{a_{CB}}(t)=-\langle \phi_B(t)\vert \nabla V\vert \phi_C(t)\rangle+c.c.$ and the CC transition as $\mf{a_{CC}}(t)=-\langle \phi_C(t)\vert \nabla V\vert \phi_C(t)\rangle$.

The saddle-point approximation (valid for $\omega\ll I_p \ll U_p$) is successively applied to both expressions. This way, the momentum integration is carried out first with saddle points defined via $\nabla_{\mf{p}}S=0$ yielding $p_s(t,t')=-\int_{t'}^{t}d\tau A(\tau)/(t-t')$ for a field polarized linearly along the x-axis.
The time integration is more involved as the ionization matrix elements exhibit a singularity at the saddle points of the $t'$ (and $t''$) integration defined by $\partial_{t'} S(p_s(t,t'),t,t')-I_p=0$  but can be treated in a similar way as in Ref.~\cite{CHIRILA2002}. Finally, we obtain
\begin{eqnarray}
 a_{CB}(t)&=&-\text{Re}\frac{(8 I_p)^{5/4}}{4 }\sum_{t'} \sqrt{n_0(t)n_0(t')}w_{corr}(t') D(t,t')\nonumber\\
&\times&\langle 0\vert \partial_x V\vert p_s(t,t')+A(t)\rangle   e^{-i(S(p_s(t,t'),t,t')+I_p(t- t'))} \label{a_0c4}\nonumber\\
a_{CC}(t)&=&\text{Re}\frac{(8 I_p)^{5/2}}{32} \sqrt{n_0(t')n_0(t'')}w_{corr}(t')w_{corr}(t'') \nonumber\\
&\times&D(t,t')D(t,t'')\, \langle p_s(t,t')\vert \partial_x V\vert p_s(t,t'')\rangle \nonumber\\
&\times&e^{-i(S(p_s(t,t''),t,t'')-S(p_s(t,t'),t,t')+I_p(t'-t''))}\nonumber\label{a_cc4},
\end{eqnarray} 
where $D(t,t')=\sqrt{(2 \pi i)^3/(t-t')^3}/(d^2/dt'^2S(p_s(t,t'),t,t'))$, $w_{corr}(t)=\sqrt{w(t)/w_{K}(t)}$ and $t'<t''$.
The ionization correction factor $w_{corr}(t)$  was inserted~\cite{IVANOV1996} as the non-Coulomb corrected SFA contains only the Keldysh ionization rate $w_{K}(t)$ which is not sufficient in this regime.
The matrix elements $\langle \mf{p}\vert \partial_x V\vert\mf{p}'\rangle$ and $\langle 0\vert \partial_x V\vert\mf{p}\rangle$ can be calculated analytically for a spherically symmetric Coulomb potential  $V(r)=-\sqrt{\alpha}/r$ with the ground state wave function $\psi_{1s}=\alpha^{3/4}/\sqrt{\pi}e^{-\sqrt{\alpha}r}$ where $\alpha=2 I_p$ which yields 
 $\langle p_x\vert -\partial_x V\vert q_x\rangle=\frac{i \sqrt{\alpha}}{2\pi^2} \frac{1}{p_x-q_x}$ and
 $\langle p_x\vert -\partial_x V\vert 0\rangle= i\sqrt{2} \alpha^{5/4} \big(p_x -\arctan(p_x/\alpha)\big)/(\pi p_x^2)$.

\begin{figure}
\centering
\includegraphics[width=\linewidth]{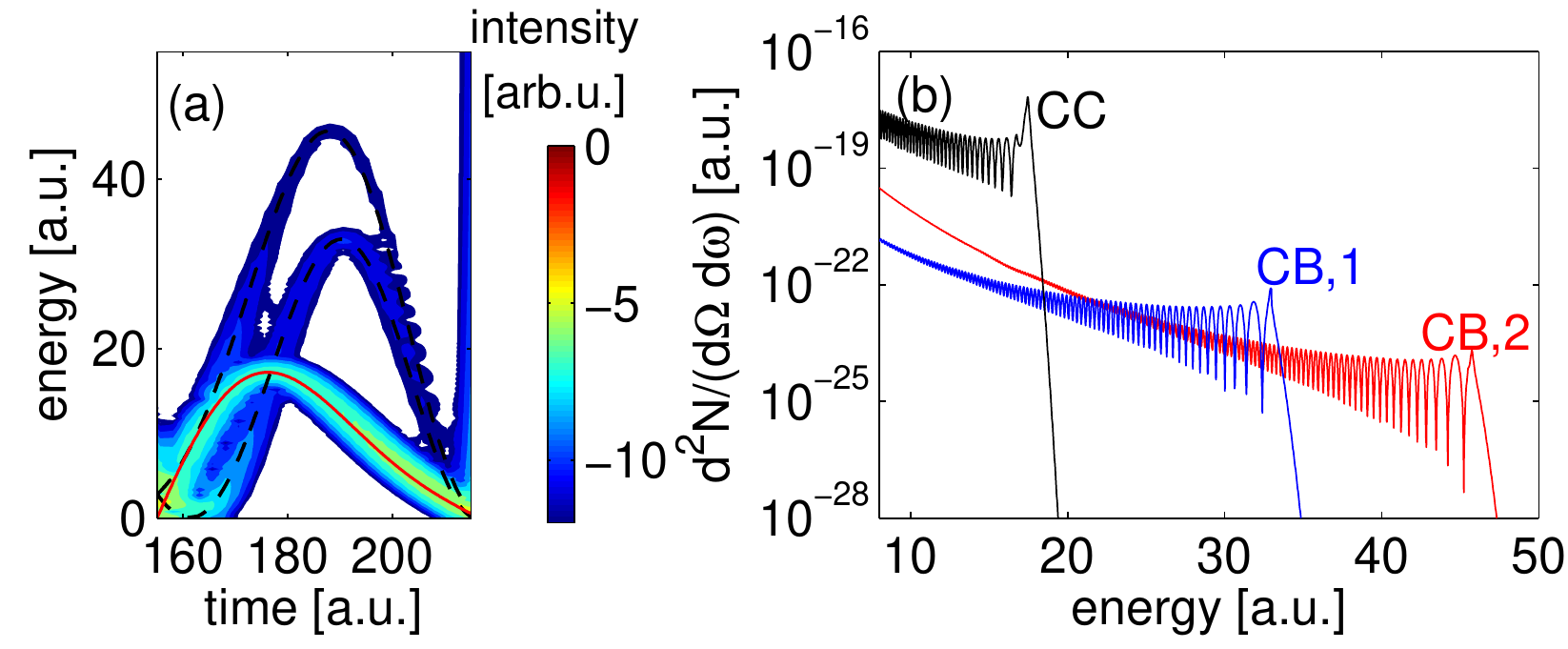}
\caption{\label{fig-HHGstrongfield}
(color online) CC HHG by using an analytical approach based on the SFA for the same field as in Fig.~\ref{fig-HHGnumerical}. (a) is the windowed Fourier transform of the acceleration within a chosen time window in excellent agreement with the TDSE results of Fig.\ref{fig-HHGnumerical}. (b) shows single-atom spectral photon yield per solid angle per shot of the same time window as in (a). }
\end{figure}

The analytically calculated dipole acceleration is shown in Fig.~\ref{fig-HHGstrongfield}, for the same electric field and in the same manner as for Fig.~\ref{fig-HHGnumerical}.
The excellent agreement with the TDSE results shows that the process is correctly described by the SFA model.  The SFA model allows to extract the phase of the emitted harmonics $\phi=S(p_s(t,t''),t,t'')-S(p_s(t,t'),t,t')+I_p(t'-t'')-\omega_H t$ converted to $\phi(I)=\alpha_{cc}I$ as a function of pulse peak intensity $I$, which reveals a striking difference for CB as compared to CC transitions. For the time window considered in Fig. \ref{fig-HHGstrongfield}, we find 6 contributions to the harmonic energy at $\omega_H=8$ a.u.: Two CB transitions from the wave packet ionized at the first peak of the laser pulse ($\alpha_{CB,1,s}=+0.6\times 10^{-13}$cm$^2$/W and $\alpha_{CB,1,l}=+1.5\times 10^{-13}$cm$^2$/W for the short and long trajectory, respectively), two CB transitions from ionization at the second peak ($\alpha_{CB,2,s}\rightarrow 0$ and $\alpha_{CB,2,l}=+1.4\times 10^{-13}$cm$^2$/W) and the two CC transitions ($\alpha_{CC,s}=-0.6\times 10^{-13}$cm$^2$/W and $\alpha_{CC,l}=-0.1\times 10^{-13}$cm$^2$/W). As a result, and as can be understood by the difference-energy mechanism, the sign of $\alpha$ for CC vs. CB harmonics is found to be different and thus allows to experimentally isolate the CC harmonics.

The impact of our results could be manifold: 
First of all, CC HHG may qualitatively advance and complete tomographic molecular imaging: Instead of sensing the orbital shape of the active electron~\cite{ITATANI2004}, the effective potential could be measured. This is because the detailed structure of the potential enters the CC transition mainly via the matrix element $\langle p'\vert \nabla V\vert p''\rangle$ which is the Fourier transform of the gradient of the effective potential, which is not restricted to an ionic potential.  
 This could be particularly interesting for larger molecules, where the entire molecular potential, i.e. mainly the positions of all nuclei, would be mapped out, not just the few highest-lying molecular orbitals.

Moreover, it is known that an electron ionized from one isolated atom cannot coherently recombine with another atom to produce CB HHG, as there is no phase relation among isolated atoms.  However, the phase of the CC transition does not depend on any bound-state portion of a wavefunction, only on the shape of the potential. It would  allow for coherent interferences in another atomic or more complex potential even if they are some fixed distance away.

Also, CC transitions contribute to harmonic energies below $I_p$ on which currently a frequent discussion exists about their underlying mechanism~\cite{HU2002}.

The authors would like to thank K. Z. Hatsagortsyan and D. Bauer for fruitful discussions. Financial support from the MPRG program of the Max-Planck-Gesellschaft is gratefully acknowledged (C.O., P.R., R.H., I.S., T.P.).


\end{document}